\pdfoutput=1
\documentclass{sig-alternate-2013}

\usepackage{color}
\usepackage{multirow}
\usepackage{amsmath,lipsum}
\usepackage{amsmath}
\usepackage{nameref}
\usepackage{url}

\usepackage[
  pass,
]{geometry}

\newcommand{\newcomment}[3]{%
  \expandafter\newcommand\csname #1\endcsname[1]{\textcolor{#2}{[#3: ##1]}}%
}

\newcomment{bill}{blue}{Bill}
\newcomment{jevin}{red}{Jevin}
\newcomment{poshen}{cyan}{Poshen}

\permission{Copyright is held by the authors}
\conferenceinfo{}{} 
\copyrightetc{}
\crdata{}

\begin{document}


%

\title{Viziometrics: Analyzing Visual Information in the Scientific Literature}

%
%
%
%
%

\numberofauthors{3} 
%
\author{
%
%
\alignauthor
Po-shen Lee\\
       \affaddr{University of Washington}\\
       \affaddr{185 Stevens Way}\\
       \affaddr{Seattle, Washington 98105}\\
       \email{sephon@uw.edu}
\alignauthor
Jevin D. West\\
       \affaddr{University of Washington}\\
       \affaddr{Box 352840}\\
       \affaddr{Seattle, Washington 98195}\\
       \email{jevinw@uw.edu}\textbf{}
\alignauthor 
Bill Howe\\
       \affaddr{University of Washington}\\
       \affaddr{185 Stevens Way}\\
       \affaddr{Seattle, Washington 98105}\\
       \email{billhowe@cs.washington.edu}
}

\date{30 July 1999}

\maketitle
\begin{abstract}
Scientific results are communicated visually in the literature through diagrams, visualizations, and photographs. These information-dense objects have been largely ignored in bibliometrics and scientometrics studies when compared to citations and text. In this paper, we use techniques from computer vision and machine learning to classify more than 8 million figures from PubMed into 5 figure types and study the resulting patterns of visual information as they relate to impact. We find that the distribution of figures and figure types in the literature has remained relatively constant over time, but can vary widely across field and topic.  Remarkably, we find a significant correlation between scientific impact and the use of visual information, where higher impact papers tend to include more diagrams, and to a lesser extent more plots and photographs.  To explore these results and other ways of extracting this visual information, we have built a visual browser to illustrate the concept and explore design alternatives for supporting viziometric analysis and organizing visual information. We use these results to articulate a new research agenda -- viziometrics -- to study the organization and presentation of visual information in the scientific literature.
\end{abstract}

%
%

%
%



\section{Introduction}
Significant information in the scientific literature is conveyed visually using plots, photographs, illustrations, diagrams, and tables.
This information is crafted for human consumption and, unlike the surrounding text, is not directly machine-readable. This lack of programmatic access has led to relatively few studies exploring how these visual encodings are used to convey scientific information in different fields and how patterns of encodings relate to impact.

The visual cortex is the highest-bandwidth information channel into the human brain \cite{ware12} and humans are known to better retain information presented visually~\cite{nelson:76}.  The figures in the scientific literature therefore would appear to play a critical role in scientific communication.  The discovery of the structure of DNA was largely a visual argument based on the images produced by X-ray crystallography; indeed, Gibbons argues that the act of producing the visualization of the structure represents the discovery itself~\cite{gibbons:12}.  The first extra-solar optical images of planets amplified the nascent subfield of astronomy focused on planet-hunting~\cite{kalas:08}.  Medical imagery of biological processes at scales below that which can be be detected using conventional optical methods are providing new insight into brain function~\cite{dani:10}.
In all fields, key experimental results are summarized in plots, complex scientific concepts are illustrated schematically in diagrams, and photographic evidence are used to provide insight at scales and in locations not available to the human eye.  

In the 1950s, researchers like Eugene Garfield and De Solla Price recognized the importance of citations in organizing and searching the scientific literature~\cite{garfield:06,desollaprice:65}, but the process for making this information useful at scale was painstaking. We see an analogy with the current role of the visual literature.  There is clear value in extracting and analyzing this information to understand its role on scientific communication and impact, just as there is clear value in analyzing the citation network. The citation network tells us how ideas are related; visual representations tell us how ideas are communicated. Figures from related groups, authors, and fields share a `DNA' that can reveal how information is conveyed.

We adopt the term \emph{viziometrics} to describe this line of research to convey the shared goals with bibilometrics and scientometrics.  In this paper, we present an initial exploration of viziometrics by analyzing a corpus of papers from PubMed to relate the use and distribution of visual information with impact, and consider how these patterns change over time and across fields. Specifically, we consider three questions:

\begin{itemize}
\item How do patterns of encoding visual information in the literature vary across disciplines?
\item How have patterns of encoding visual information in the literature evolved over time?
\item Is there any link between patterns of encoding visual information and scientific impact?
\end{itemize}

To answer these questions, we built a platform called VizioMetrix for exploring the visual literature. VizioMetrix includes components for ingesting a corpus of papers, a database for managing the extracted metadata, analysis routines for dismantling multi-chart images, a classifier for identifying figure types, and a public figure-oriented search and browse interface that illustrates a different approach to organizing the scientific literature in terms of visual results and concepts rather than the papers that contain them.

Our key result is a link between the use of scientific diagrams (schematics, illustrations) and the impact of the paper, suggesting that high-impact ideas tend to be conveyed visually. We conjecture two possible explanations for this link: that visual information improves clarity of the paper, leading to more citations and higher impact, or that high-impact papers naturally tend to include new, complex ideas that require visual explanation.  More broadly, we argue that identification and description of the visual patterns, verified through computational experiments spanning a large corpus of papers, can help improve understanding of how scientific information is best conveyed, how the organization of visual information relates to scientific impact, how to present scientific information more accessibly to a broader audience, and perhaps most directly, how to build better services for organizing, browsing, and searching the ``visual literature.''

\def\arraystretch{2.0}
\begin{table}[tb]
\centering
\caption{We classified 4,781,741 figures into six categories. The table shows the number of figures for each figure type before and after dismantling.}

\medskip
\label{table:fig_stats}
\begin{tabular}{l|p{2cm}p{2cm}}
\multirow{2}{*}{Figure Type} & Count Before Dismantling & Count After Dismantling\\
 \hline\hline
Multi-chart & 1,416,237 (29.6\%) & None\\
\hline
Equation &1,425,042 (29.8\%)  & 1,741,059 (17.0\%)\\
Diagram & 652,918 (13.7\%) & 2,036,704 (19.9\%)\\
Photo & 475,615 (9.9\%) & 2,322,231 (22.7\%)\\
Plot & 475,327 (9.9\%) & 3,579,839 (35.0\%)\\
Table & 336,602 (7.1\%)& 553,171 (5.4\%)\\
\hline
Total & 4,781,741 & 10,233,004\\

\end{tabular}
\end{table}

\def\arraystretch{2.0}
\begin{table}[tb]
\centering
\caption{Evaluation of multi-chart figure classifier and figure-type classifier using 10-fold cross validation.}
\medskip
\label{table:recall_precision}
\begin{tabular}{l|c|c}
Figure Type & Precision & Recall \\
 \hline\hline
Multi-chart & 92.9\% & 86.3\%\\
Singleton & 89.3\% & 94.6\% \\
\hline
Equation & 95.4\% & 95.1\% \\
Diagram & 84.2\% & 84.1\%\\
Photo & 94.5\% & 97.3\%\\
Plot & 91.5\% & 90.2\% \\
Table & 95.1\% & 93.1\% \\

\end{tabular}
\end{table}

\section{Related Work}

Computer vision techniques have been used in the context of conventional information retrieval tasks (retrieving papers based on keyword search), including some commercial systems such as D8taplex \cite{D8taplex} and Zanran \cite{Zanran}. Search results from these proprietary systems have not been evaluated and do not appear to  make significant use of the semantics of the images. 

In 2001, Murphy et al. proposed a Structured Literature Image Finder (SLIF) system, targeting microscope images \cite{murphy2001searching}. A decade later, Ahmed et al. ~\cite{Ahmed2010,ahmed2009structured} improved the model for mining captioned figures. The latest version combines text-mining and image processing to extract structured information from biomedical literature. The algorithm first extracts images and their captions from papers, then classifies the images into six classes. Classification information and other metadata can be accessed via web service. However, SLIF focuses exclusively on microscropy images and does not extend to general figures.

Choudhury et al. \cite{ray2015architecture} proposed a modular architecture to mine and analyze data-driven visualizations that included (1) an extractor to separate figures, captions, and mentions from PDF documents \cite{choudhury2013figure}, (2) a search engine \cite{bhatia2010finding}, (3) raw-data extractor for line charts \cite{kataria2008automatic, browuer2008segregating, lu2007, ChoudhuryaGREC15}, and (4) a natural language processing module to understand the semantics of the figure. Also, they presented an integrated system from data extraction to search engine for user experience. Chen et al. \cite{Chen2015} proposed their search engine named DiagramFlyer for data-driven figures. It recovers the semantics of text components in the statistical graph. Users can search figures by giving attributes of axes or the scale range in further. Additionally, DiagramFlyer can expand queries to include related figures in terms of their production pipelines. 

Although these early projects represent a different approach for information retrieval tasks, they make no attempt to analyze the patterns of visual information in the literature longitudinally.  In this paper, we present our figure processing pipeline that classifies figure images in different categories (Section: \nameref{sec:data_method}) and a search interface that uses these classified images as the primary unit of interaction to facilitate search tasks (Section: \nameref{sec:ui}). The key result of this paper involves the analysis of the figures in a large collection of publicly available papers (Section: \nameref{sec:result}).

\begin{figure*}
 \centering
 \includegraphics[width=6.5in]{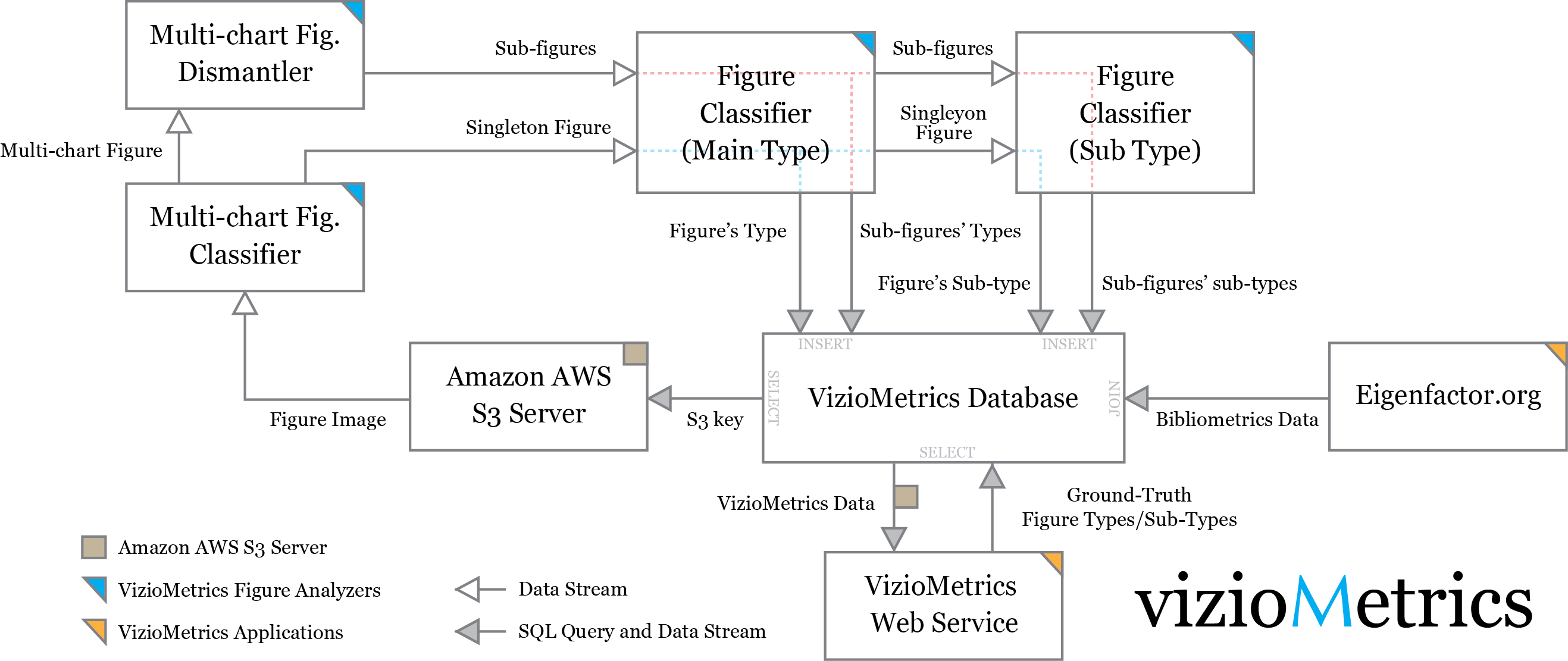}
 \caption{VizioMetrix system overview. We store the images in Amazon's S3 service. Image paths, figure captions, paper metadata and classification result are stored in the database. The figure analysis system acquires the file keys from the database, downloads the image files, and feeds them into the figure processing pipeline. The final classification results are stored in the database as the sources for the application prototype.}
 \label{fig:flowchart}
\end{figure*}

\section{Dataset and Methodology}
\label{sec:data_method}
We developed a platform called VizioMetrix \footnote{We distinguish the Vizio\textbf{M}etrix platform from the field of study (\textit{Viziometrics})} with which we analyzed 4.8 million figures from more than 650,000 PubMed Central (PMC) papers (7.4 paper).  PubMed Central, an archive of biomedical and life science literature, provides free access to the full text documents including the source images.  We downloaded the article files from the PMC FTP server and extracted the images into a figure corpus.  Of these files, about 66\% had figure files. These figure files are separated from the PDF files so we can skip the step of extracting them from literature. In addition, PMC also provides paper metadata including paper titles, authors, publishing data, citations, and image captions that we use in our figure search engine and analysis by field. 

We found five image formats in use: GIF, JPEG, TIF, TIFF, PNG.  The vast majority (99\%) of the images were in JPEG format with a small number of PNG files.  We had several filtering steps to remove duplicate images and the images that are not scientific figures. First, we removed all GIF files since they are duplicates of images in other formats. Second, we removed image files that are prints of full papers. Third, we converted all TIF and TIFF files to JPEG files and resized their dimension such that the longer edge was 1280 pixels.  We did not modify the aspect ratios if the original image size was larger than this value. 

After filtering, we classified 4.8 million images into five categories.  The classification algorithm is described in Section: \nameref{sec:figure_analysis}.  The classifier returns a probability distribution across all class types, but for each image we only assigned the label with the highest probability.  The class labels are as follows:

\begin{itemize}

\item  Equation (e.g., embedded equations, Greek and Latin characters)
\item  Diagram (e.g., schematics, conceptual diagrams, flow charts, architecture diagrams, illustrations) 
\item  Photo (e.g., microscopy images, diagnostic images, radiology images, fluorescence imaging)
\item  Table (any tabular structures with text or numeric data in the cells)
\item  Plot (e.g., bar charts, scatter plots, line charts)  

\end{itemize}    

There were 1.4 million figures that contain multiple sub-figures within one image (for example, with sub-figures labeled part A, part B, etc.) which we refer to as \emph{multi-chart} figures.  We ``dismantled'' these multi-chart figures into their individual parts using a custom algorithm that we developed for this purpose \cite{Lee2015}.  After dismantling, we extracted and classified another 5 million individual figures.  In total, we classified more than 10 million figures.  

The results of our classification are summarized in Table \ref{table:fig_stats}. This summary information alone provides some interesting insights:  About 67\% of the figures are embedded in multi-chart figures, and plots are the most likely figure type to be embedded in this way: we found 475k standalone plots but 3.5M total plots after dismantling.  Tables are significantly less common than other figure types suggesting a preference among authors (or possibly editors) for presenting results visually.  There is a relatively uniform distribution across diagrams, photos, and plots; the prevalence of photos is likely an artifact of the biomedical emphasis of the PMC corpus. 

\begin{figure*}[tb]
 \centering
 \includegraphics[width=6.5in]{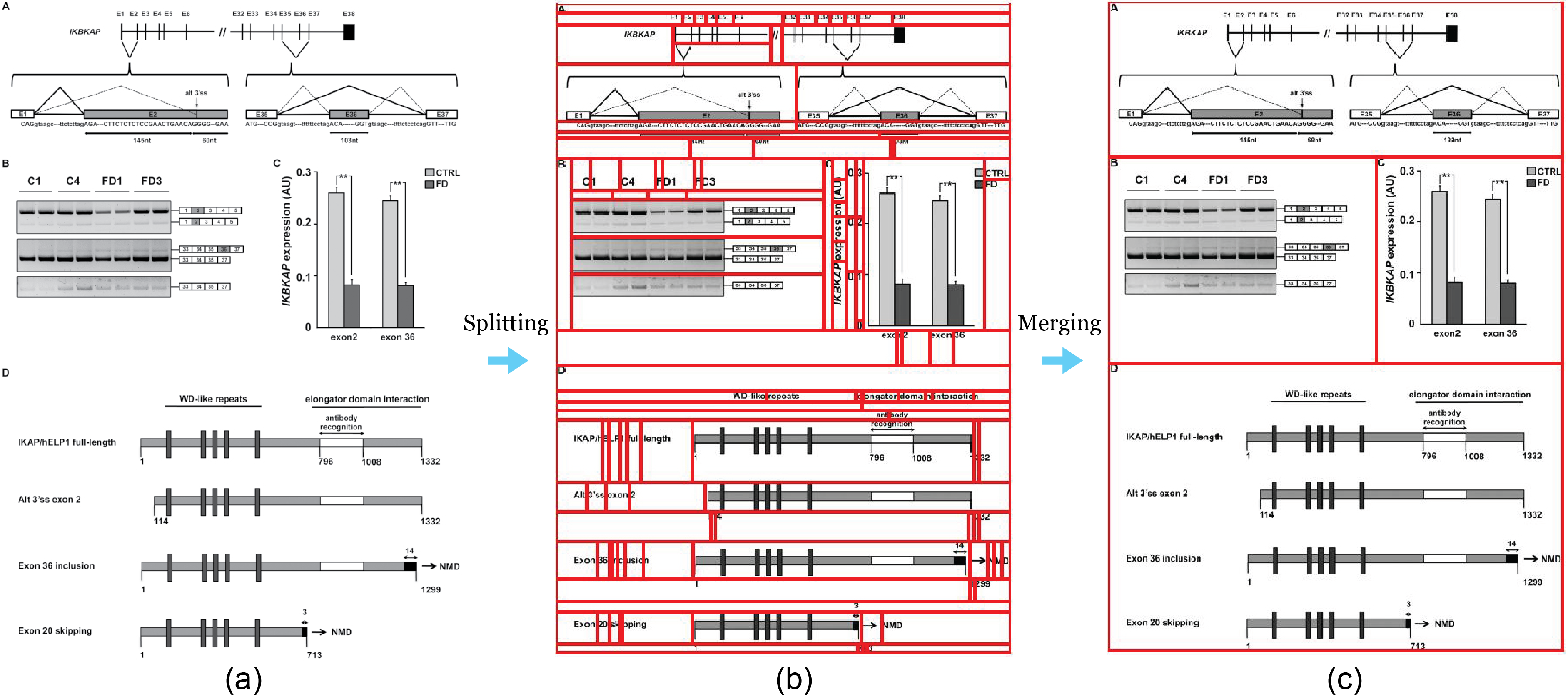}
 \caption{Multi-chart figure dismantling.  The figure shows the intermediate steps for dismantling multi-chart figures. The splitting algorithm recursively segments the raw images into several sub-images. The merging algorithm then aggregates auxiliary fragments with nearby standalone figures to produce the final segmentation.}
 \label{fig:dismantle}
\end{figure*}

\subsection{Figure Analysis}
\label{sec:figure_analysis}
Figure \ref{fig:flowchart} illustrates the analysis pipeline used to perform classification.  We first download and extract the images in AWS (Amazon Web Services).  We then classify each figure as either \emph{multi-chart} or \emph{singleton}.  Each figure identified as multi-chart is dismantled into a set of singleton figures.  All singleton figures (including those dismantled from multi-chart figures) are labeled with one of five class labels: equation, diagram, photo, plot and table.  The classified images can be browsed online at \url{viziometrics.org}. In the following sections, we will briefly describe the algorithm for each box in Figure \ref{fig:flowchart}.  

\subsubsection{Figure Classification} 
\label{sec:figure_classification}

The primary technique we used to classify the images is called ``Bag of Features,'' a computer vision approach that extracts small patches (as features) from images, uses them to build a codebook, and then re-encodes the images using a histogram. We adopt the technique developed by Coates et al. \cite{coates2011analysis} and extended by Savva et al. \cite{Savva2011}. 

First, we normalize an image to a 128$\times$128 grayscale image with a constant aspect ratio.  Then, we randomly extract a set of 6$\times$6 patches from each training image and normalize the contrast of each patch. To reduce the cross-correlation between patches, PCA whitening \footnote{Adjacent pixel values can be highly correlated. Whitening makes patches less correlated with each other.} is applied on the entire patch set. Next, we run k-means on the patches with k = 200 to identify 200 common patch types, one for each cluster. A representative patch for each patch type, called a \emph{codebook}, is derived from each cluster. For each training image, we generate a new set of patches by sliding a 6$\times$6 window in one-pixel increments across the image.  For each such generated window patch, we find the most similar codebook patch (via Euclidean distance) and increment a counter for that codebook.  The set of codebook counters forms a histogram, and this histogram forms the feature vector used to train the classifier.  To account for the global structure of common visualizations  (e.g., axes are typically found on the left and bottom of the image), each image is split into four quadrants and a separate 200-element histogram is computed for each quadrant.  The final feature vector of 800 elements is obtained by concatenating the four 200-element histograms. These feature vectors are then classified using a Support Vector Machine (SVM).

We implemented this approach in Python 2.7 using scikit-learn's implementation of SVMs. The corpus we used for training was randomly sampled from the PMC corpus (\url{ftp://pub/pmc/ee/}).  We manually labeled 3,271 images as one of five categories: photos (782), tables (436), equations (394), visualizations (890), and diagrams (769) and used these hand-labeled data to train the classifier. For each category, we randomly reserved 25\% of the images for a testing set and trained the SVM model on the remaining 75\%. We used sklearn's grid search method to tune the SVM parameters (kernel, gamma, and penalty parameter)\footnote{The optimized model is run by the RBF kernel with gamma of 0.001 and penalty parameter of 1000.}. Once the model parameters are tuned, we evaluate the model by using the testing set and then trained the final model with all images. In this paper, we report the evaluation of classification performance (Table \ref{table:recall_precision}) by 10-fold cross-validation on the full training corpus of 3,271 images. The final classification accuracy for all images is 91.5\% 

\begin{figure*}[tb]
 \centering
 \includegraphics[width=6.5in]{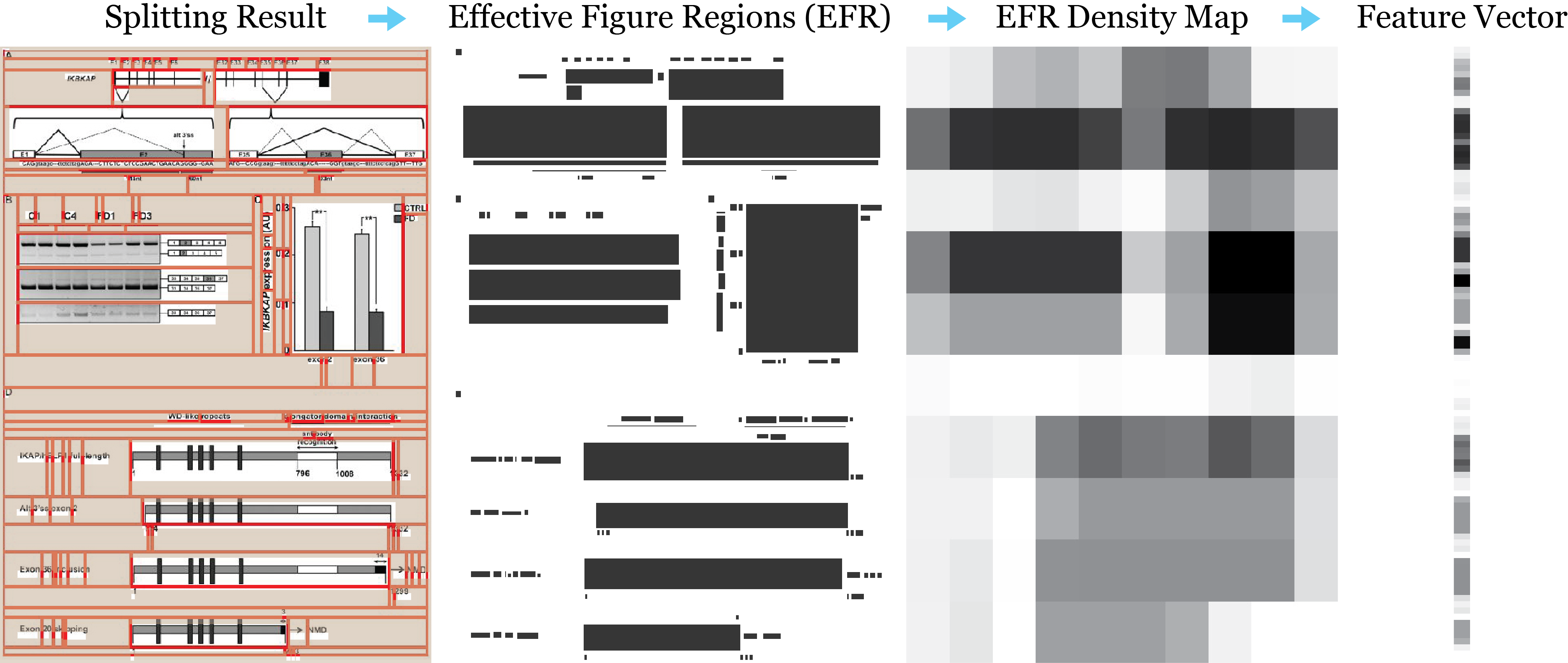}
 \caption{Recognizing mulfti-chart images.  After splitting the figure into distinct blocks, the dismantling algorithm marks the effective figure regions (EFR) then downsamples the EFR into $n\times n$ blocks that form a $n^2 \times 1$ feature vector. These vectors are used to train the classifier. }
 \label{fig:detector}
\end{figure*}

\subsubsection{Figure Dismantling}
\label{sec:figure_dismantling}
The use of multi-chart figures complicates classification. For example, the figure in Figure \ref{fig:dismantle} consists of four sub-figures: a photo, a plot, and two diagrams. Approximately 30\% of all figures in our corpus required dismantling. We designed a \emph{dismantling} algorithm to extract the component figures. Our algorithm first splits an image into fragments based on background color and layout patterns. An SVM-based binary classifier then distinguishes complete charts from incomplete auxiliary fragments (e.g., axis labels, tick marks, and legends). Next, we recursively merge fragments to reconstruct complete charts, choosing between alternative merge trees using a scoring function based on heuristics. Figure \ref{fig:dismantle} illustrates the process of splitting a source image then merging the fragments to produce a final output. The algorithm correctly extracted 82.9\%\footnote{The algorithm produced 2743 sub-images, where 2499 of them are considered correct. From manually dismantling, we found 3013 sub-figures in the 500 multi-chart figures. The recall is 82.9\% and the precision is 84.3\%} of the sub-images from 500 multi-chart figures randomly sampled from PMC. The details of the algorithm are explained in an earlier paper \cite{Lee2015}.

\subsubsection{Multi-chart Figure Classification}
\label{sec:multi_classification}
Attempting to dismantle every figure image in our corpus would be prohibitively expensive and extremely wasteful; only about 30\% of the images are multi-chart figures. We therefore developed a simple and fast pre-classifier to distinguish multi-chart figures from singleton figures in order to reduce the number of dismantled singletons. 

We designed the method based on two observations: that multi-chart figures tend to have a different size and shape than singleton figures, and that the layout of a multi-chart figure tends to follow a regular grid pattern. Based on these two observations, we constructed a feature vector with K (K = M + N) elements: M elements based on the size and shape, and N elements based on the grid layout. The M elements consist of the image height ratio($height_{i}$ / $height_{avg}$) and the image width ratio($width_{i}$ / $width_{avg}$) where the denominators are average image height and average image width of all images in the training set respectively. The N elements are derived from the output of splitting algorithm of the dismantler. 

Figure \ref{fig:detector} shows the splitting, and the red lines indicate the boundaries between fragments. For each block, we mark the minimal rectangular region that contains non-empty pixels, so that we can obtain the effective figure regions (EFR) and use them as a mask. We subdivide the mask into $n\times n$ blocks and compute the proportion of EFR in each block as defined as the EFR density map. Finally, we squeeze the values into a 1-D vector with $n^2$ elements. 

We set $n = 10$ as the final parameter (M = 100) and apply the same technique described in the previous section to train the figure classifier.  The final model is optimized by using a RBF kernel with gamma of 0.001 and a penalty parameter of 1000. We obtained 91.8\% accuracy by 10-fold cross-validation on the entire training set comprising 880 multi-chart figures and 1067 singleton figures. The recall and precision for each class are shown in Table \ref{table:recall_precision}. 

\subsection{Measuring Scholarly Influence}
\label{sec:measure_scholar}

To assess the influence of a particular paper, we used the article-level Eigenfactor (ALEF) score \cite{west2016ALEF}.  ALEF is a modified version of the PageRank algorithm \cite{ilprints422} that captures a random walk on the paper-level citation graph where each vertex is a paper and each directed edge is a citation.  Because a random walker will move inexorably backwards in time using the standard PageRank approach, we modify the algorithm to correct for this.  The modified algorithm reduces the number of steps the random walker takes and teleports the random walker to links rather than nodes \cite{west2016ALEF, lambiotte2012ranking}.  

The ALEF ranking method has been shown to outperform simple citation counts and standard PageRank approaches \cite{WesleySmith2016wsdm}.  The ALEF method took second place in a recent data challenge sponsored by the ACM International Conference on Web Search and Data Mining (WSDM) \footnote{http://www.wsdm-conference.org/2016/wsdm-cup.html}. Although ALEF is the state-of-the-art measure of impact that has been shown to provide better static rankings of scientific papers \cite{WesleySmith2016wsdm}, the qualitative results of this study would not change if we simply used citation counts as our measure of impact.

\begin{figure}
 \includegraphics[width=3.3in]{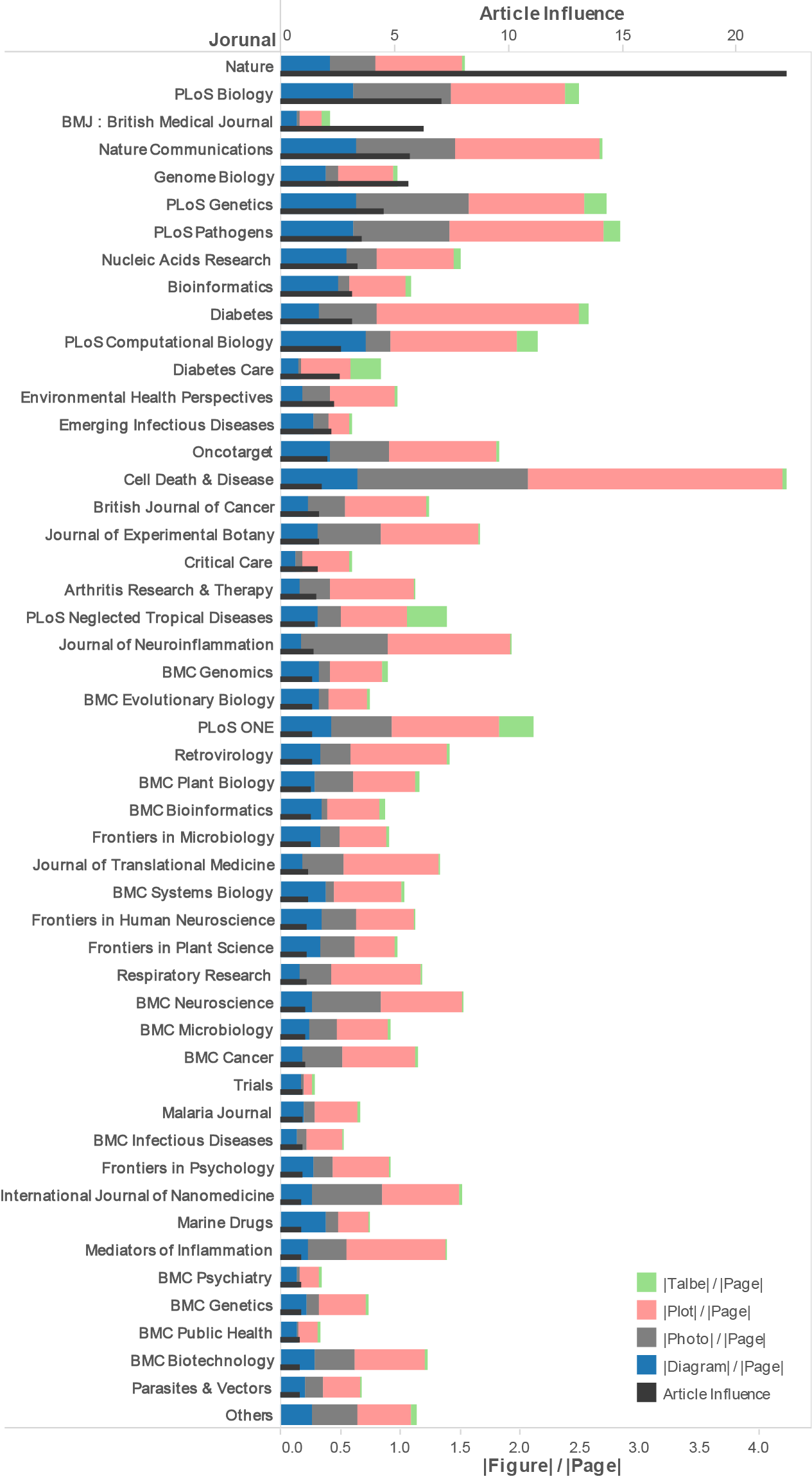}
 \caption{The distribution of figure types across journals show an emphasis on plots and diagrams relative to tables, and identify visualization-heavy venues such as Cell Death and Disease.  We considered the top 49 highest-impact journals in PMC that had at least 850 papers available in the corpus, where impact is measured as Article Influence (AI) (the black bar).  Each stacked bar shows the average density of each figure type across all papers published in the journal.  The density of a figure type is the number of instances of that type divided by the page count.   The category ``Others'' contains 288,953 papers from other journals.}
 \label{fig:journal_AI}
\end{figure}

\section{Exploring Visual Patterns in the Literature}
\label{sec:result}

We use the classified figures to study patterns in the use of visual information across scientific domains, across publication venues, and over time.  We also used the classifications to examine the effect on scholarly impact.  Our key result is a correlation between the influence of a paper and the distribution of visualizations: A higher proportion of diagrams corresponds to a higher impact paper.  To verify these results, we analyze a number of potential sources of statistical error in the methods and data and show that the results are robust (see supplementary material).   

In a broader sense, we are interested in better understanding how complex results are and should be communicated to other scientists and to the general public.  There is a widening technical gap that can be closed through effective visualizations. We see an opportunity in training students on how best practices.  In this paper, we focus on diagrams and schematics.  We find that higher impact (higher cited) papers tend to have a higher concentrations of diagrams and schematics.  The next step will be to look for more subtle patterns of effective visualization in diagrams.  For example, what kinds of diagrams are most effective?  What features of diagrams make the easy or less easy to convey?  Now that we can classify them at large scale, we can build datasets for testing this.

\subsection{Data Details}
\label{sec:data_detail}

We use the images described in Section \nameref{sec:data_method} with further refinements for visual pattern analysis. We only include papers for which ALEF scores exist for which and page numbers can be determined, since we need these metrics in our calculations of impact.   We also remove papers published before 1997 since the annual quantity is too low to produce meaningful results  (less than 300 for each year). 

After these three steps, our set includes 494,663 papers and 6,897,810 figures after dismantling, excluding equations.  We exclude equations because it is difficult to assess the number of equations in a figure. Some of the PMC literature is in pre-print formats rather than the official journal format.  For these papers, we use the total number of pages from PMC. As a result, the page count can be different from their official copies. In addition, we underestimate the total number of tables from those authors who use only latex or Microsoft word to build tables, since these authors typically do not provide tables as separate images. 

The dataset does not necessarily represent all of scholarship. Authors of the papers analyzed here can voluntarily select to submit papers to PMC, and PMC will clearly tend to attract papers in the life sciences with an emphasis on human biology.  In particular, \textit{Nature} publishes a significant number of Physics papers, but these papers will be underrepresented in PMC. 

\subsection{Visual Patterns Across Disciplines}
\label{sec:pattern_disciplines}

To analyze the patterns of visual encodings across disciplines, we normalize the individual figure counts by the total number of pages in order to measure the density of each figure types.  This figure count normalization is similar to the method used by Fawcett et al. \cite{fawcett2012heavy} in their analysis of equations.  It ensures the values are comparable between articles with diverse lengths. 

\begin{figure}
 \includegraphics[width=3.3in]{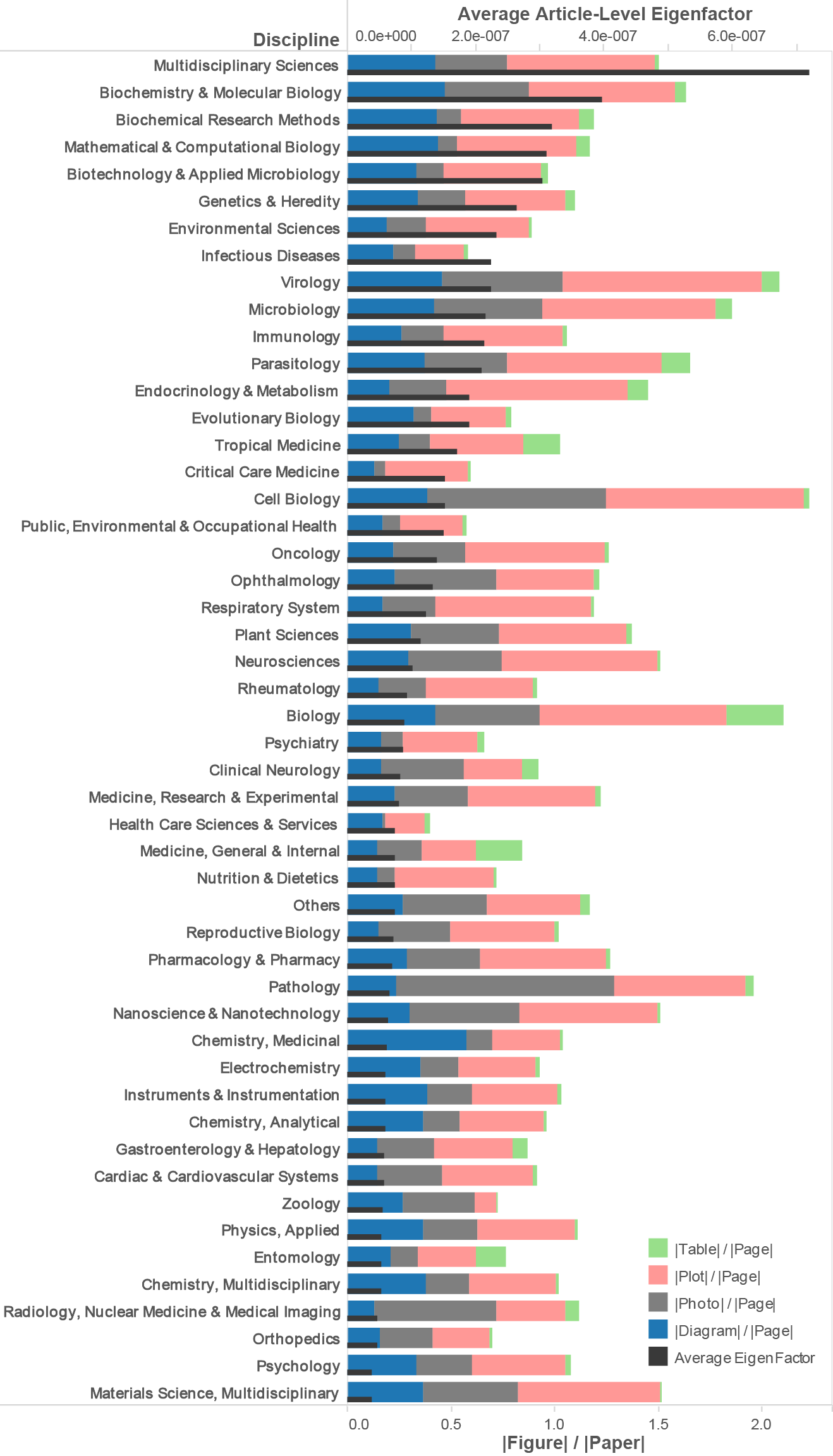}
 \caption{Figure distribution by research topic show that microbiology topics tend to emphasize visual presentation of ideas. Topics were determined by the journal categories in Thomson Reuters' JCR.  We show the highest-impact 49 topics that have at least 1000 papers, where impact is the average of all papers assigned to that category. The category ``Others'' includes 216,380 papers from other topics and papers without topic labels.}
 \label{fig:discipline_EF}
\end{figure}

Next we aggregate the figures and papers by journal and research topics to see how figure types vary across publishing venues and disciplines. Figure \ref{fig:journal_AI} and Figure \ref{fig:discipline_EF} show the average figure density of journals and research topics for which we were able to collect at least 850 or 1000 papers published during 1997 to 2014 from PMC, respectively.  Figure topics were assigned used Thomson-Reuters' Journal Citation Report (JCR) category system.  

In order to include Nature in the diagram, we set the threshold to 850 articles per journal. The stacked bars present the densities of diagrams, photos, visualizations, and tables from the left to the right. Equations are not considered in this case because defining the quantity of equations can be vague: a single image may contain any number of equations. In Figure \ref{fig:journal_AI}, the thin dark bars display the impact of the journals, as measured by ArticleInfluence (AI) for the journal \cite{west2010eigenfactor}. In Figure \ref{fig:discipline_EF}, we used the average ALEF score to estimate the value of topic areas because topic areas consist of overlapping journals. The AI scores is a citation metric for measuring journal influence \cite{west2010eigenfactor}.  The underlying citation data comes from Thomson-Reuters' JCR.  Journals and research topics are listed by impact in descending order. Due to the limit of page capacity, we show only the top 49 items and gather the papers from small-collection journals and lower-rank journals into ``Others.'' 

Figure \ref{fig:journal_AI} shows the top 49 journals ordered by AI.  Differences exist between journals.  The journal \textit{Cell Death and Disease} relies heavily on microscopy and experimental evidence, and we see this emphasis manifest as a significantly higher number photos and plots. We find that multidisciplinary journals, such as the \textit{Nature} series and the \textit{PLoS} series exhibit a balance of figure types. Qualitatively, many of the journals with high figure-per-page counts are also high in AI. Further, papers from the top one-third journals (16 out of 50) tend to have more diagrams. Journals emphasizing case studies are exceptions: \textit{British Medical Journal}, \textit{Diabetes Care}, and \textit{Emerging Infectious Diseases}.  In comparison, papers from the journals near the tail show lower diagram density. We will make this observation statistically precise in Section: \nameref{sec:pattern_impact}.

Using Thomson Reuters' JCR, we can assign each journal to a research topic, then repeat the analysis of figure distribution by research topic rather than journal. We describe the method used to assign topic labels in more detail in Section: \nameref{sec:measure_scholar}.


Figure \ref{fig:discipline_EF} shows the disciplines for which at least 1000 papers were available. Differences in figure type density exist between disciplines.  For example, \textit{cell biology} and \textit{pathology} have a relatively high number of photos per page, whereas \textit{mathematical and computational biology} and \textit{medicinal chemistry} have fewer photos per page and relatively more diagrams and plots per page.  \textit{Biology} and \textit{internal medicine} tend to have relatively more tables per page, suggesting an emphasis on (or tolerance of) presenting quantitative results numerically.  We conjecture that these patterns have more to do with cultural norms for publication rather than specific research methods, but we have not studied these questions in this preliminary effort. 

\subsection{Visual Patterns Over Time}
\label{sec:pattern_over_time}

\begin{figure}[tb]
 \includegraphics[width=3.3in]{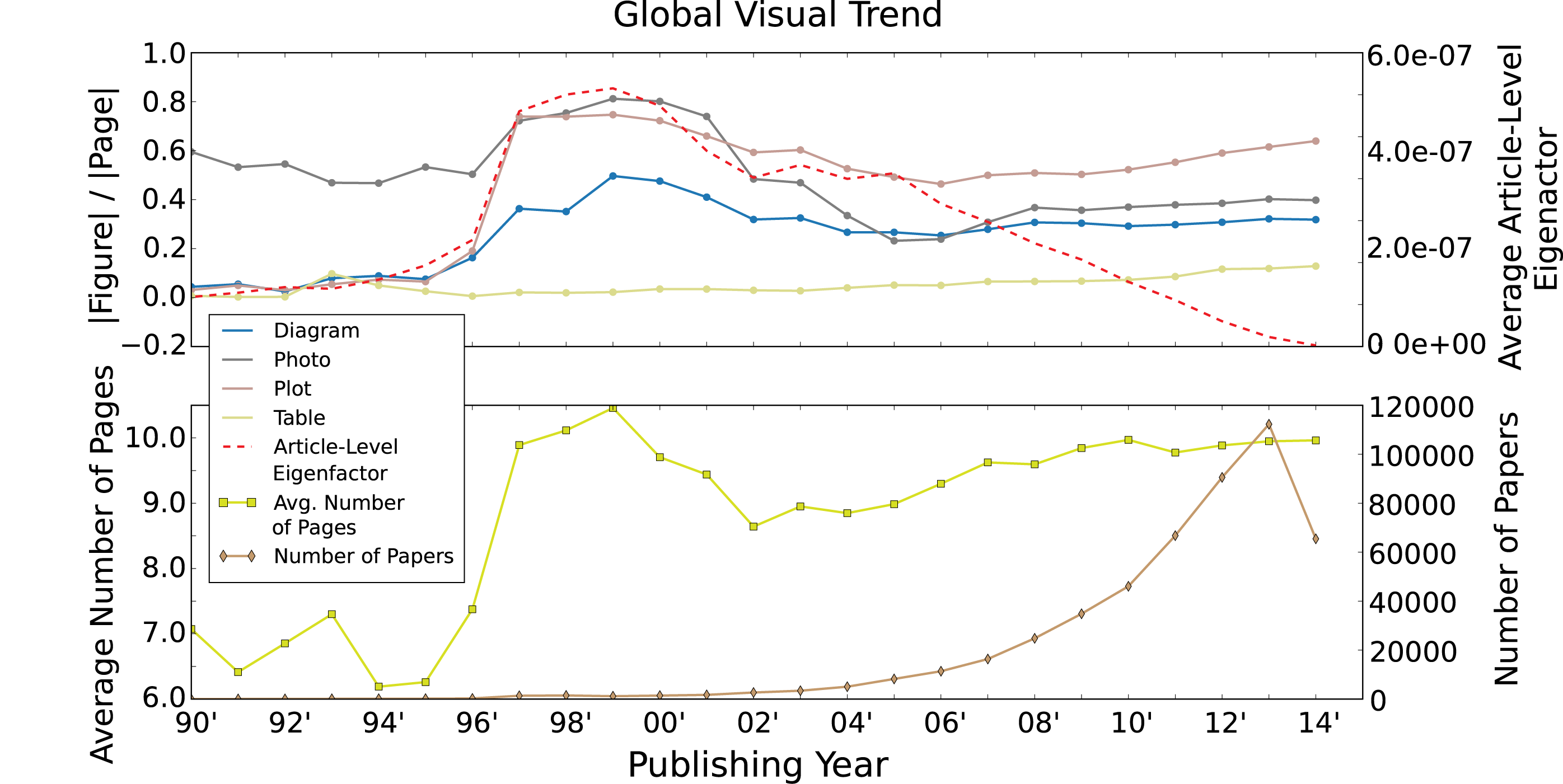}
 \caption{The distribution of figure types in the PMC corpus over time.  The top figure shows the number of papers increasing dramatically in the mid-2000s, which can be explained by a change in sponsor rules: NIH required authors to submit their papers to PMC. The "hump" of impact between 1997 and 2005 may be attributable to author bias in voluntarily uploading their highest-impact papers. After 2006, the increasing uses of plots and tables may be attributable to increased emphasis on data-intensive research. The density of photos and diagrams are consistently flat over time.  The bottom plot provides context: the average page length per paper over time, and the number of papers in the corpus over time.}
 \label{fig:2_1}
\end{figure}

We analyze patterns of visual information over time by segmenting the data into different publishing years. The earliest paper we collected from PMC was published in 1937, but relatively few papers earlier than 1997 are included (biasing the corpus).  We plot the total number of papers in our database from 1990 to 2014 in Figure \ref{fig:2_1}. Paper quantity reaches the thousand mark in 1997 and the ten thousand mark in 2007.  In 2008, NIH mandated that authors upload their papers to PMC, partially explaining the growth of the corpus. Papers can be uploaded at any time for any publication year, so we do not necessarily see an increase in later papers. The average ALEF score increases until 2000 and then decreases, consistent with most measure of impact that are inherently time-sensitive. 


The "hump" that occurs in Figure \ref{fig:2_1} around 1997 to 2002 is attributable to a bias in the corpus; in this period, the corpus was dominated by just three journals: \textit{Journal of Cell Biology}(38\%), \textit{Journal of Experimental Medicine}(31\%), and \textit{Journal of General Physiology}(8\%). As more journals were added to PMC, this sampling bias decreased, and the patterns stabilized.   After 2006, the number of diagrams per page remains relatively consistent, and a small but consistent growth in the number of plots and tables per page is observed.  We conjecture that these increases are attributable to an increased emphasis on data-intensive science in the biological and biomedical disciplines. 


\begin{figure}[tb]
 \includegraphics[width=3.3in]{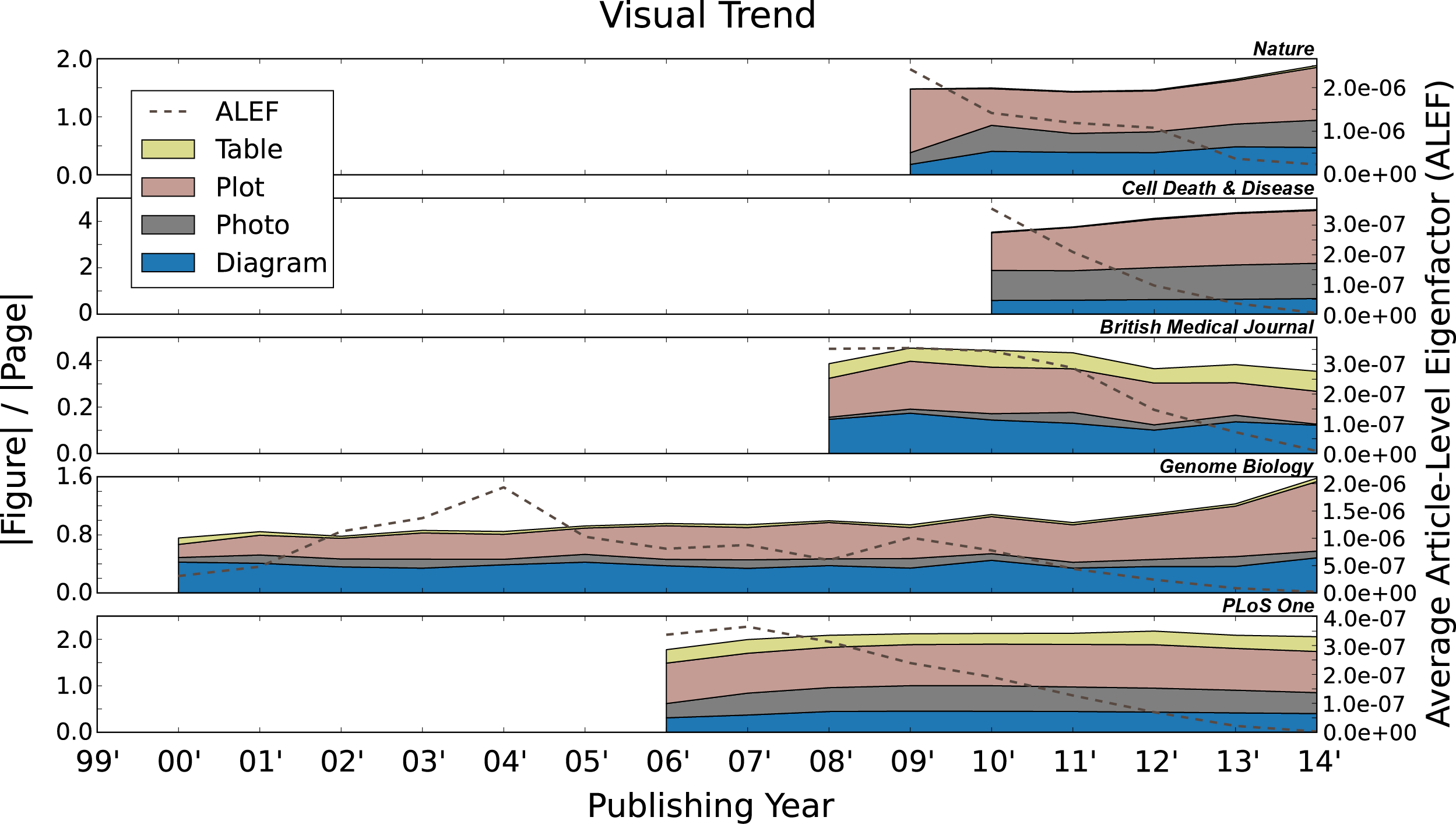}
 \caption{We choose five specific journals for closer inspection: \textit{Nature} (highest impact), \textit{Cell Death and Disease} (highest figure density), \textit{British Medical Journal} (lowest figure density), \textit{Genome Biology} (unusually low proportion of photos) and \textit{PLoS One} (largest number of papers). \textit{Nature, Cell Death and Disease} and \textit{Genome Biology} exhibit a recent increase in plots-per-page, consistent with the overall trend. We conjecture that the articles in these high-impact journals are becoming more data-centric.  Moreover, \textit{Nature} and especially \textit{Cell Death and Disease} show a heavy use of figures, in part because these journals tend to have greater proportions of multi-chart figures (67\% for \textit{Nature} and 82\% for \textit{Cell Death and Disease} relative to 30\% for the entire image set.) The \textit{British Medical Journal} shows a different trend in which figure density gradually decreases; the mechanism behind this trend is unclear. \textit{PLoS One} shows no significant change from its launch in 2006.}
 \label{fig:2_2}
\end{figure}



In Figure \ref{fig:2_2}, we select five journals with unique features for closer inspection: \textit{Nature} (highest impact according to our measures), \textit{Cell Death and Disease} (highest figure density), \textit{British Medical Journal} (lowest figure density), \textit{Genome Biology} (unusually low proportion of photos) and \textit{PLoS One} (largest number of papers). \textit{Nature} exhibits an increase in figure density over time, driven primarily by an increase in plot density which may reflect an increased emphasis in data-intensive science.  For the journal \textit{Cell Death and Disease}, one sees the same effect of growing figure density over time, which corresponds to an increased use of multi-chart figures: 81\% of the figures are multi-chart compared to an average of 38\%.\footnote{Equations are not taken into account.} In contrast, the \textit{British Medical Journal} exhibits low figure density and a gradual decrease in the use of figures over time.  Tables are used more in proportion compared to most journals and photos are extremely rare.  We conjecture that the decrease in visual information over time may be related to a known shift in focus for BMJ, in which the editor has intentionally focused on topics of broad public interest \cite{bmj2016}.  It is possible that heavy use of quantitative data in the form of plots may make articles \emph{less} accessible.  \textit{Genomics Biology} was selected for its unusually low proportion of photos, which appears consistent over time.  We do see the density of plots increasing significantly since 2011, following the global trend. We selected \textit{PLoS One} because of the extremely large number of papers in the corpus. Because it is broadly multidisciplinary, the patterns of figures represent many fields of study and we do not expect, nor do we see, any distinctive pattern.  \textit{PLoS One} may represent a microcosm of the overall literature in this regard.

\begin{figure}[tb]
 \includegraphics[width=3.3in]{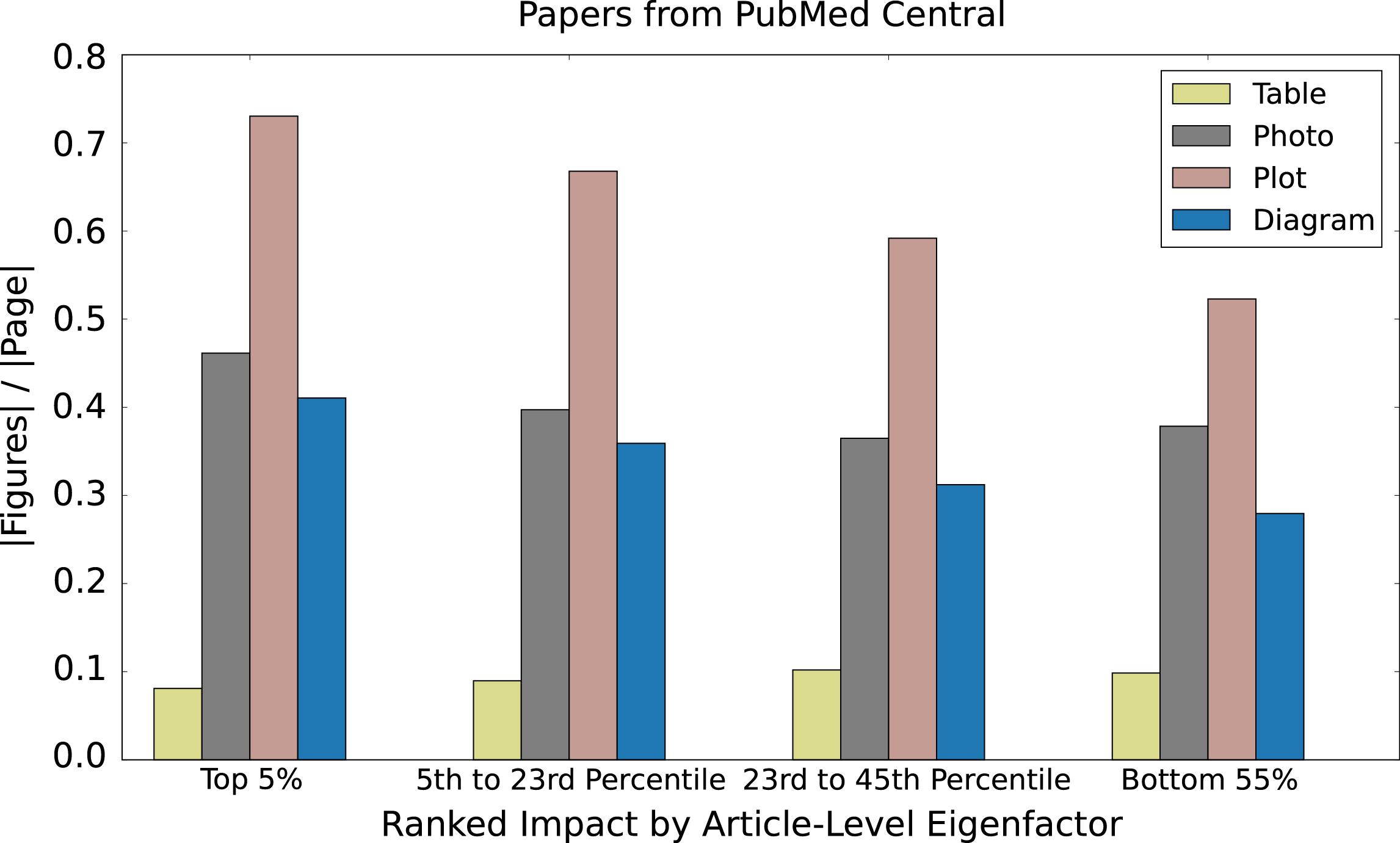}
 \caption{Impact versus figure density.  We rank papers by ALEF and group them into 4 bins. Papers with the same Eigenfactor are grouped into the same set. Any two papers with Eigenfactor difference within 1E-12 are regarded as having the same impact, which is why the bins are not evenly distributed. For each set, we average the densities of 4 figure types. There are statistically significant correlations for plot (0.099570 +/- 0.000027) and diagram (0.110295 +/- 0.000032).}
 \label{fig:3_0}
\end{figure}

\begin{figure}[tb]
 \includegraphics[width=3.3in]{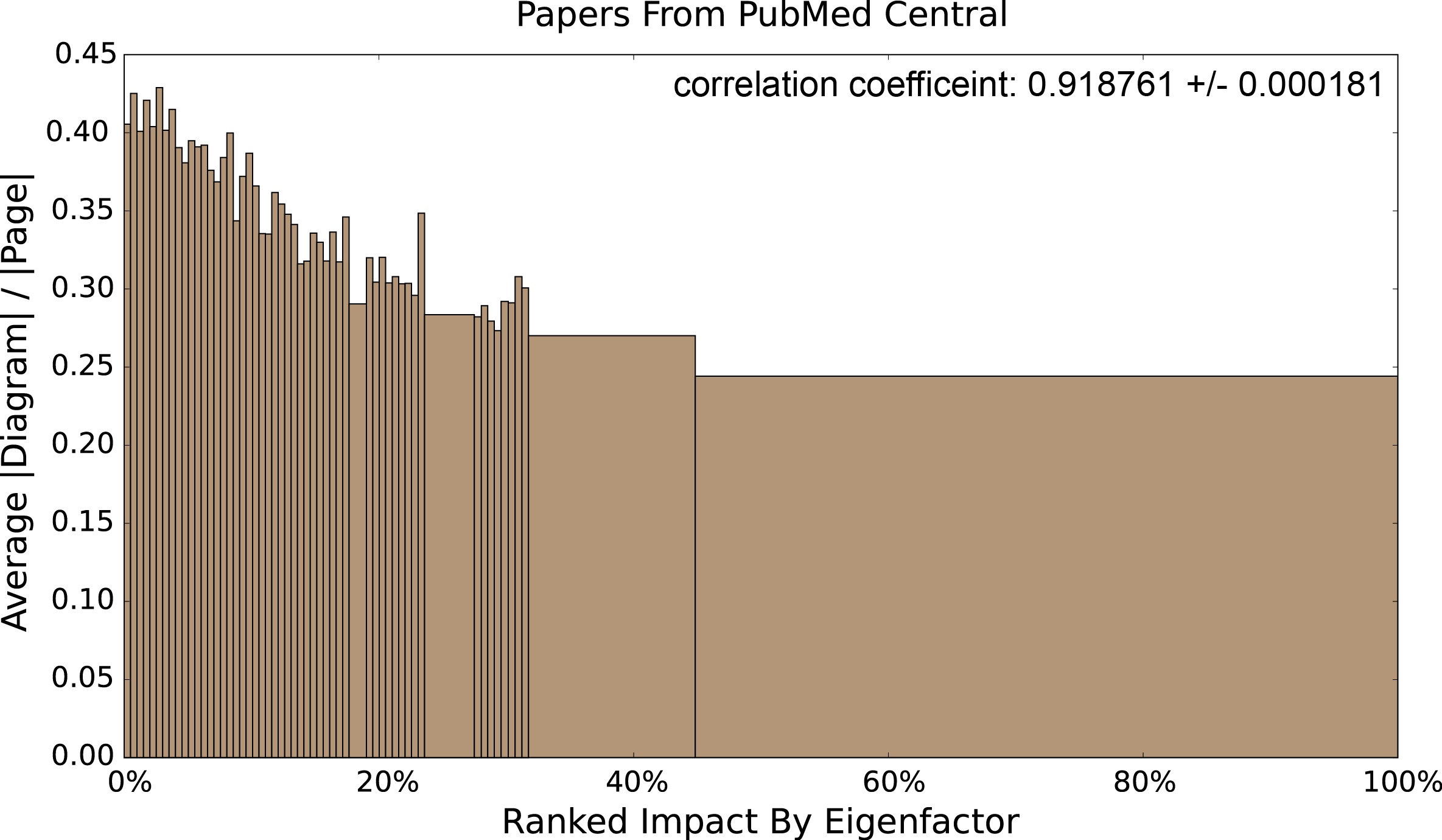}
 \caption{Considering \textit{PLoS One} accounts for 21.5\% papers of the total, we filter them to eliminate the \textit{PLoS One} bias and reproduce the histogram from the rest 392,992 papers. Each bin counts for 1980 papers to achieve a fraction of 0.5\%. Finally 56 bins are produced. A stronger binned correlation (0.92) is obtained than the binned correlation acquired from using the full dataset (0.84).
 }
 \label{fig:56bin_no_plos_one}
\end{figure}

\def\arraystretch{2.0}
\begin{table}[]
\scriptsize
\centering
\caption{We estimate the correlation between the ALEF score and figure density (left column) and proportion of figures (right column). Each table entry $X(Y)$ indicates the correlation including ($X$) and excluding ($Y$) papers from \textit{PLoS One}, a journal that tends to bias the results due to a high proportion of tables.  Correlations excluding \textit{PLoS One} are more strongly positive for all figure types. The entry NSS indicates that the result was not statistically significant. Overall, high proportions of diagrams are linked to high impact while high proportions of photographs are linked to lower impact (negative correlation).} 
\medskip

\label{table:coeff}
\begin{tabular}{l|p{2cm}|p{2cm}}
\multirow{3}{*}{Figure Type} &  \multicolumn{2}{c}{Correlation Coefficient} \\
\cline{2-3}
		   & Figure Density\newline (w/o PLoS One) & Prop. of Figure  \newline(w/o PLoS One)\\
 \hline\hline
 
Diagram & 0.84 (0.92)  & 0.61 (0.52)\\ 
Photo & 0.57 (0.70) & -0.69 (-0.63)\\
Plot & 0.60 (0.80) & NSS (NSS)\\ 
Table & NSS (0.78) & NSS (NSS) \\
\end{tabular}
\end{table}

\subsection{Visual Patterns Related to Impact}
\label{sec:pattern_impact}

In this section, we consider the relationship between patterns of visual encodings and scientific impact.    


Figure \ref{fig:3_0} shows qualitatively that higher impact papers tend to have a higher density of plots, diagrams and photos and a lower density of tables suggesting that the use of visual information correlates with impact. We chose four bins that characterize the Eigenfactor score distribution, which tends to follow a power law distribution.  We chose the four bins to roughly correspond to boundaries at 95\%, 75\%, 50\%.  The bin boundaries are not these numbers exactly because many papers have identical Eigenfactor scores,\footnote{Any two papers with Eigenfactor difference within 1E-12 are regarded as having the same score.} and we did not want to artificially separate two papers with the same score into two different bins. Instead, we move the boundary to the next highest threshold.  The bin boundaries then become 5\%, 23\%, and 45\%, with the lowest bin (Bottom 55\%) containing all papers with Eigenfactor score of zero.  For each group, we average the figure densities for each of four figure types and produce a histogram as shown in Figure \ref{fig:3_0}. We verified that our classifiers and dismantler exhibit no bias with respect to impact (see supplementary material), indicating that the relationship we see between impact and figure type density cannot be explained by misclassification errors or dismantling errors.

The qualitative results shown in Figure \ref{fig:3_0} do not change when adjusting bin sizes. We regroup the papers binning by every half-percentile (99.5\%, 99.0\%, etc.) and compute the correlation coefficient. Table \ref{table:coeff} shows the binned correlation coefficients for the four figure types. The first and second numbers in each cell is the correlation coefficient when including and excluding papers from \textit{PLoS One} respectively. According to Figure \ref{fig:journal_AI}, \textit{PLoS One} shows a significantly higher table density then other journals, confounding the results. The key result is that higher proportions of diagrams are linked to higher impact, while higher proportions of photos are linked to lower impact (Figure \ref{fig:56bin_no_plos_one}). These results indicate that high-impact papers may tend to use more diagrams, but also that diagrams tend to be have a stronger relationship with impact than plots. One possible interpretation of these results is that clarity is paramount: illustrating an original idea visually leads to more impact then simply reporting experimental results. We conjecture that the negative correlation with photographs may suggest that tight page limits associated with high-impact journals may lead authors to sacrifice photographs as extraneous.


\begin{figure*}[tb]
 \includegraphics[width=7in]{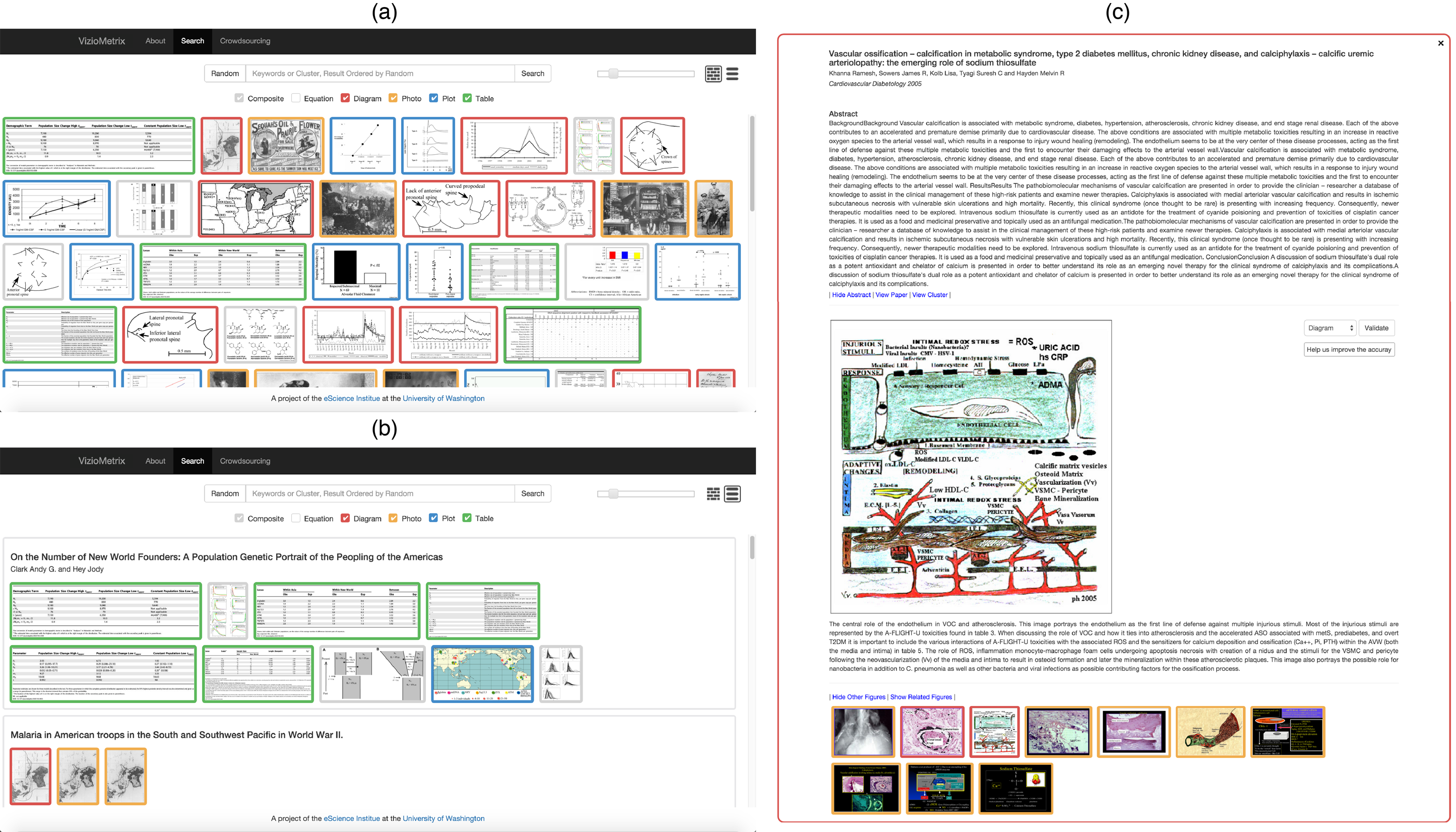}
 \caption{The user interface of the VizioMetrix search engine. Result figures are either arranged via (a) the brick-wall layout or (b) a conventional layout bundling figures with literature title. Figures are labeled by different colors based on their types. (c) Clicking figures will pop article details such as authors, abstract, figure captions, hyperlink to full PDFs and related figures. We also provide a verification form to encourage user verifying our machine-labelled figure type and help us gather more ground-truth label.
}
 \label{fig:ui}
\end{figure*}

\section{A Browser for the Visual Literature}
\label{sec:ui}

Consider a biologist who seeks a phylogenetic tree of a virus. Using a conventional academic search engine, she must enter keywords (perhaps the name of the virus and the word phylogenetic), retrieve a list of candidate papers, and, inspecting the title for relevance, open each paper for manual review.  This process operates at the wrong level of abstraction, as the search is focused on a particular method that is associated with a visual encoding (a phylogenetic tree has a distinctive visual representation).  Consider another case where a researcher wants to compare a number of different designs for solid-state laser diodes. She would like to find both scanning electron microscope (SEM) images as well as diagrams illustrating the designs, with goals of performing non-trivial analysis \emph{across} figures:  comparing the SEM photos with the corresponding diagrams (perhaps from a different paper), or a comparison of leakage currents by inspecting a set of plots showing the current-voltage curves.  With both examples, keyword search followed by manual inspection of papers to gather specific visual results seems unnecessarily inefficient.  We aim to use our classification pipeline to power a more efficient approach to this task using a figure-centric search application \cite{Lee2016www}.

The system indexes the titles, abstracts and figure captions of the corpus of papers; keyword searches probe this index to find relevant images.  Result figures are ordered by their ALEF scores, helping to reduce attention on low-impact papers. In the default layout, figures are arranged as a "brick wall" as in Figure \ref{fig:ui}(a). The color of figure border indicates its figure type as identified by our classifier.  Users can filter the figure types that are irrelevant. For instance, the biologist seeking phylogenetic trees can ignore any figures other than diagrams. For some figures with dense information such as multi-chart figures, users can use the slider to change the brick size for quick inspection, or click the figure to review the details such as title, authors, abstract, caption, related figures and more (Figure \ref{fig:ui}(c)). In addition to the brick-wall layout, we also provide conventional layout (Figure \ref{fig:ui}(b)) that lists the figures in the context of the paper in which they appear. 

\section{Future Work}
\label{sec:future}

PubMed is focused primarily in the life sciences. Future work will include extending this analysis to additional domains, enabling a comparison of visual patterns across fields of study.  We will expand our figure database with literature from diverse research areas and will continue to improve the accuracy of our classifications.  One of the key results of this paper is that more influential papers tend to have more plots and diagrams.  Next steps will be refining this question and interpreting these preliminary results to understand how figures influence impact. We plan to expand the figure processing pipeline to include additional types of figures (e.g., line charts or flow charts, or domain-specific figures such as phylogenetic trees).  

There are also many opportunities for exploring new search tools involving figure classifications.  We have received informal feedback from users on ways in which figure types could be used.  For instance, tools to support identification and directed search for specific figure types such as metabolic pathways and phylogenetic graphs could significantly accelerate research activities  In addition, information extraction from specific figure types could allow the recovery of data in support of meta-analysis activities.  Text-based search engines cannot inspect the figures and cannot  analysis and current search functions do not have the ability to extract this kind of complex information. 

One of the bottlenecks for the classifiers is the lack of labeled figures with which to train the models. We are developing a crowdsourcing component to the VizioMetrix platform that will integrate with the search service to acquire ground-truth labels as users interact with the system to complete their own tasks. The labels, images, code, and all our data will be freely available for researchers to explore their own questions.

\section{Conclusions}
\label{sec:conclusion}

In this study, we intend to launch a new field of study called viziometrics that extends  prior work in bibliometrics and scientometrics but focuses on the role of visual information encodings. We developed a figure processing pipeline that automatically classifies figures into equations, diagrams, plots, photos, and tables. By integrating the figure-type labels and article metadata, we analyzed the patterns across journals, over time, and relationships to impact.  In different disciplines, we found that the role of the five figure types can vary widely. For instance, clinical papers tend to have higher photo density and computational papers tend to have higher diagram and plot density. In respect of visual patterns over time, we found a growing use of plots, perhaps suggesting increasing emphasis on data-intensive methods.  Our key result is that high-impact papers tend to have more diagrams per page and a higher proportion of diagrams relative to other figure types. A possible interpretation is that clarity is critical for impact: illustrating an original idea may be more influential than quantitative experimental results. We also described a new application to search and browse scientific figures, potentially enabling new kinds of search tasks. The VizioMetrix systems affords search by keyword as well as figure type, and shows results in a figure-centric layout. We believe more interesting and useful applications can be inspired by the concept of viziometrics. We also encourage people to use our publicly available corpus and software to explore this area of research and create a new community of interest.

\section{Acknowledgments}
We would like to thank Dastyni Loksa for help in designing early versions of the VizoMetrix prototype. This work is sponsored in part by the National Science Foundation through S2I2 award 1216879 and IIS award III-1064505, a subcontract from the Pacific Northwest National Lab, the University of Washington eScience Institute, the Metaknowledge Network funded by the John Templeton Foundation, and an award from the Gordon and Betty Moore Foundation and the Alfred P. Sloan Foundation.

%
\bibliographystyle{abbrv}
\bibliography{jasist}  
%
%
\appendix

\section{Experiment of Verifying Statistical Significance}
\label{supl:exp_ver}

Since our classifier is imperfect, it is possible that the correlation we measure is an artifact of mistakes in classification.  One possibility is that the classifier itself behaves differently with respect to high-impact papers, perhaps making more mistakes due to the styles of figures top journals tend to attract.  Another possibility is that the errors in our classifiers just happened to produce an unusually high number of diagrams for a sufficient number of high-impact papers to generate a signal, but a perfect classifier would have shown no correlation with impact.  To test these issues, we generated two new test image sets by randomly sampling our image corpus.

To evaluate multi-chart classifier, we sampled 1000 images that are classified as singleton and 1000 images that are classified as multi-chart by the classifier and manually labeled the images for evaluation purposes.  We obtained a precision of 84.6\% from multi-chart figures and 87.3\% from singleton figures. This result is close to what we find in Table \ref{table:recall_precision}. We will use this image set to show that the error rate of the multi-chart classifier does not vary with the article impact (see Supplementary Section: \nameref{supl:classification_bias}).

The figure-type classifier may behave differently on singleton images than on multi-chart images.  In particular, we were conscious that incorrectly dismantled multi-charts could be misinterpreted as a set of diagrams, artificially inflating our estimates of diagrams, or even worse inflating more diagrams for high-impact papers. We eliminate this possibility by showing (1) the classification errors are not biased with respect to the impact and (2) the correlation results still stand when we calibrate the counts of each figure type up or down by the known error rates for the classifier. We collected a new image set for this experiment.
Considering that multi-chart figures (false-positive singletons) can be fed to the figure-type classifier in our pipeline, we randomly sampled 1400 images that are labelled as singleton from each category. Table \ref{table:confusion} shows the confusion matrix of this image set. The numbers inside parentheses denote singleton figures, while those outside the parentheses denote the sum of singleton figures and multi-chart figures of the same type (e.g. a multi-chart figure comprising 3 plots). The multi-chart figures that comprise two or more types are defined as ``composite''. These figures are false positives produced by the multi-chart figure classifier. The table shows the false singletons do inflate the number of figures particularly in diagrams. Precisely, the false diagrams mostly come from multi-plots. The low number of singleton photos is due to the failure of identifying photo arrays as multi-chart figures. These unextracted photos can be regarded as a random noise if the multi-chart classifier is not biased respect to the impact (see Supplementary Section: \nameref{supl:classification_bias}).

\begin{figure}
 \centering
 \includegraphics[width=3.3in]{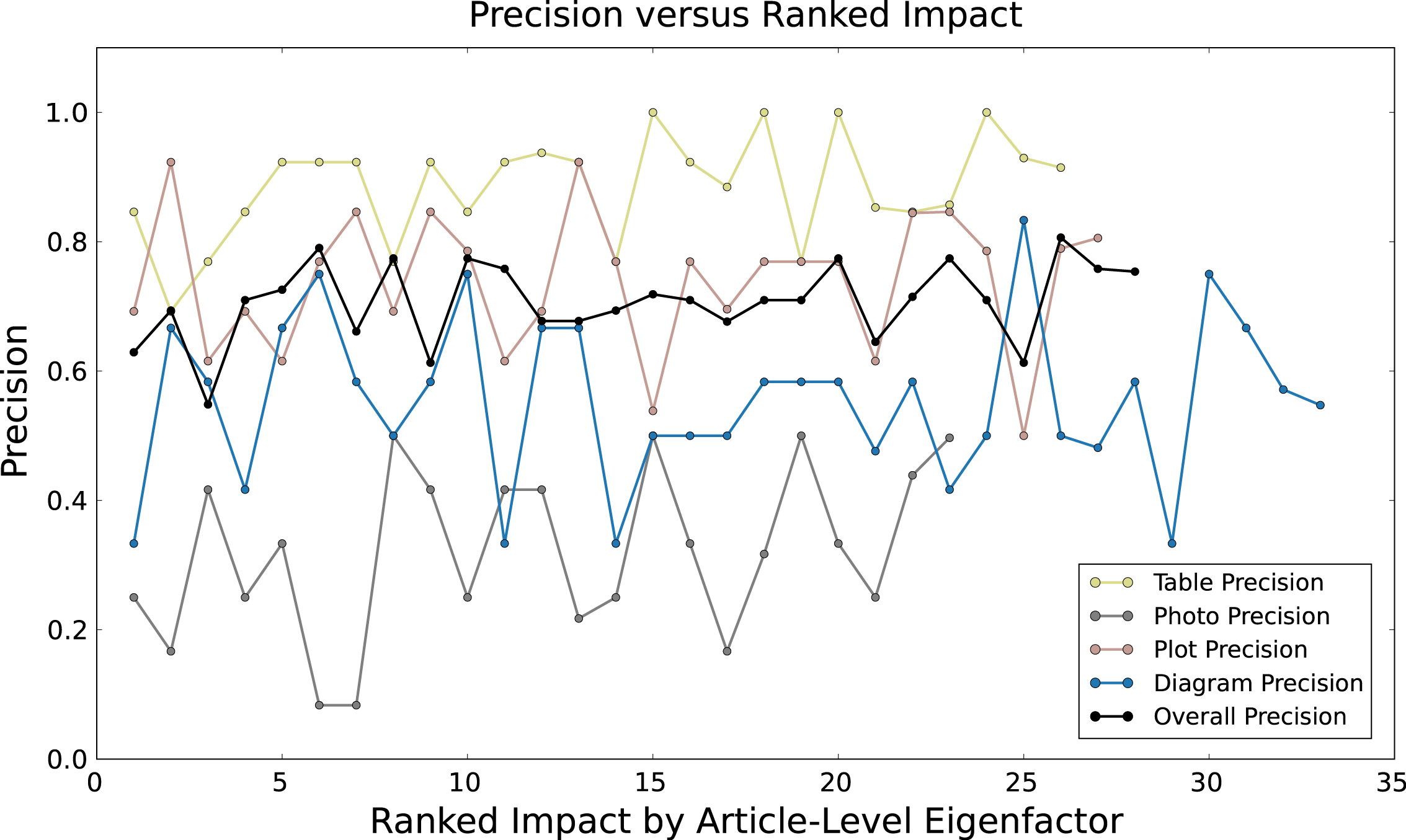}
 \caption{We randomly sampled 7000 figures (1400 for each type) that are classified as singleton and manually labelled them. To verify that the misclassification does not correlate to the article impact, we use the bin method introduced in Section: \nameref{sec:pattern_impact} to group the figures, which has been sorted by their average ALEF scores of their source papers. We filtered figures without available ALEF scores and end up with 6157 figures. Due to the small size of ground-truth data, we set the percentile to 1\%  to ensure each bin containing 10 figures or more. It shows no correlation between ``Precision All'' and the average ALEF score. Thus misclassification can be regarded as an unbiased random noise.}
 \label{fig:classifier_bias}
\end{figure}

\begin{figure}[tb]
 \centering
 \includegraphics[width=3.3in]{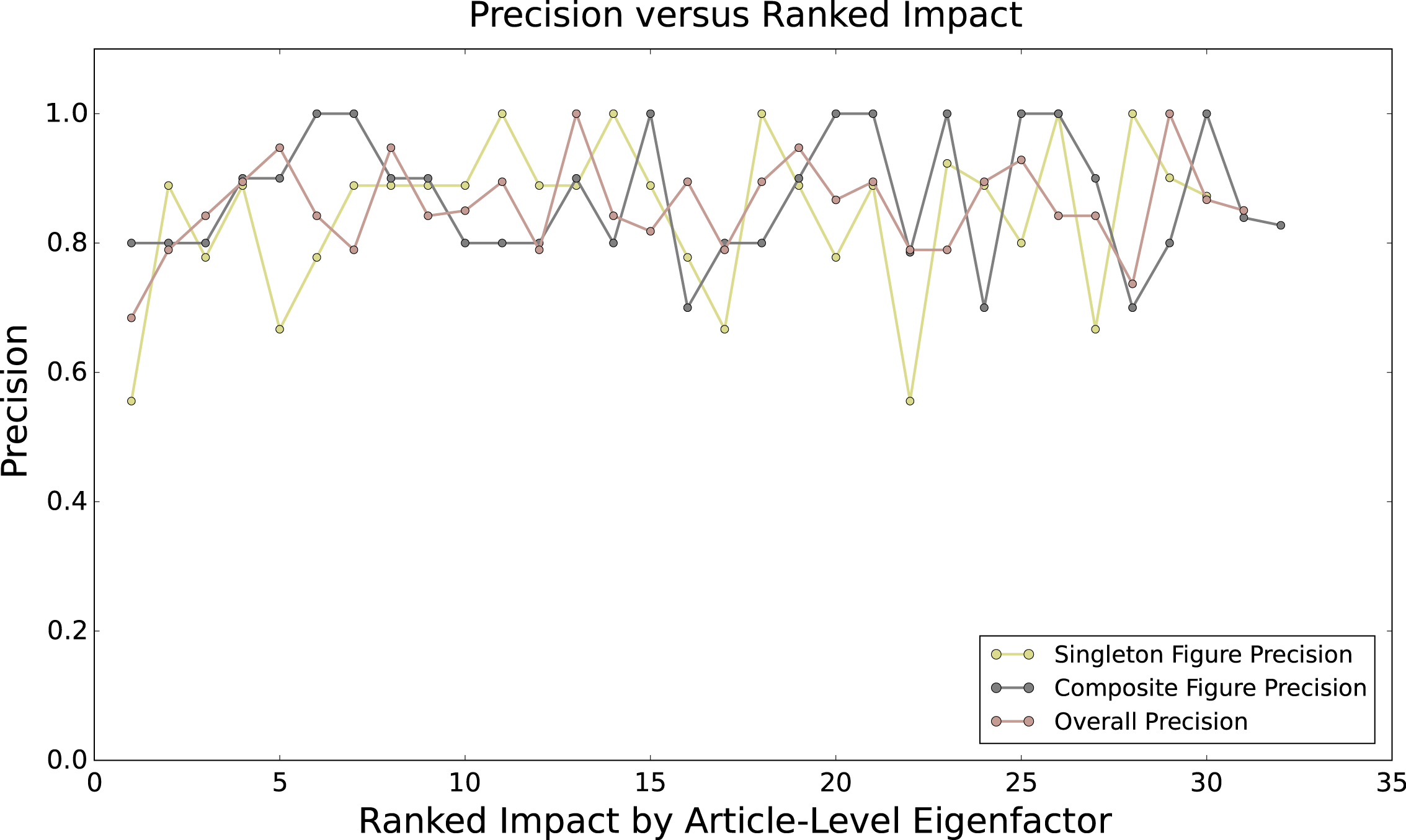}
 \caption{We randomly sampled 1000 figures that are classified as singleton and another 1000 figures that are classified as multi-chart to estimate the bias of multi-chart figure classifier. We end up with 1790 figures with available ALEF scores. By repeating the same method, it shows no correlation between the precision and the average ALEF score either.}
 \label{fig:multi-chart_classifier_bias}
\end{figure}

\subsection{Classification Bias Does Not Explain Correlation Results}
\label{supl:classification_bias}

If the classifier tended to make more mistakes on higher impact papers, the correlation estimated between article influence and figure density could be explained as an artifact of this bias.  We show that the error rate of neither of our two classifiers vary with the article impact.  
We filtered figures without available ALEF scores; from the set of testing the multi-chart classifier, we end up with 1790 figures and from the set of testing the figure-type classifier, we end up with 6157 figures. We use the bin method introduced in Section: \nameref{sec:pattern_impact} to group the figures and sort the groups by their average ALEF scores. Due to the small size of ground-truth data, we set the percentile to 1\%  and it ensures each bin containing at least 8 figures. Figure \ref{fig:multi-chart_classifier_bias} and Figure \ref{fig:classifier_bias} both show no correlation between precision and article impact. Therefore the classification errors can be regarded as an unbiased random noise. For instance, the hiding photos in the photo arrays 

\subsection{Dismantling Bias Does Not Explain Correlation Results}
\label{supl:dismantling_bias}

\begin{figure}[tb]
 \includegraphics[width=3.3in]{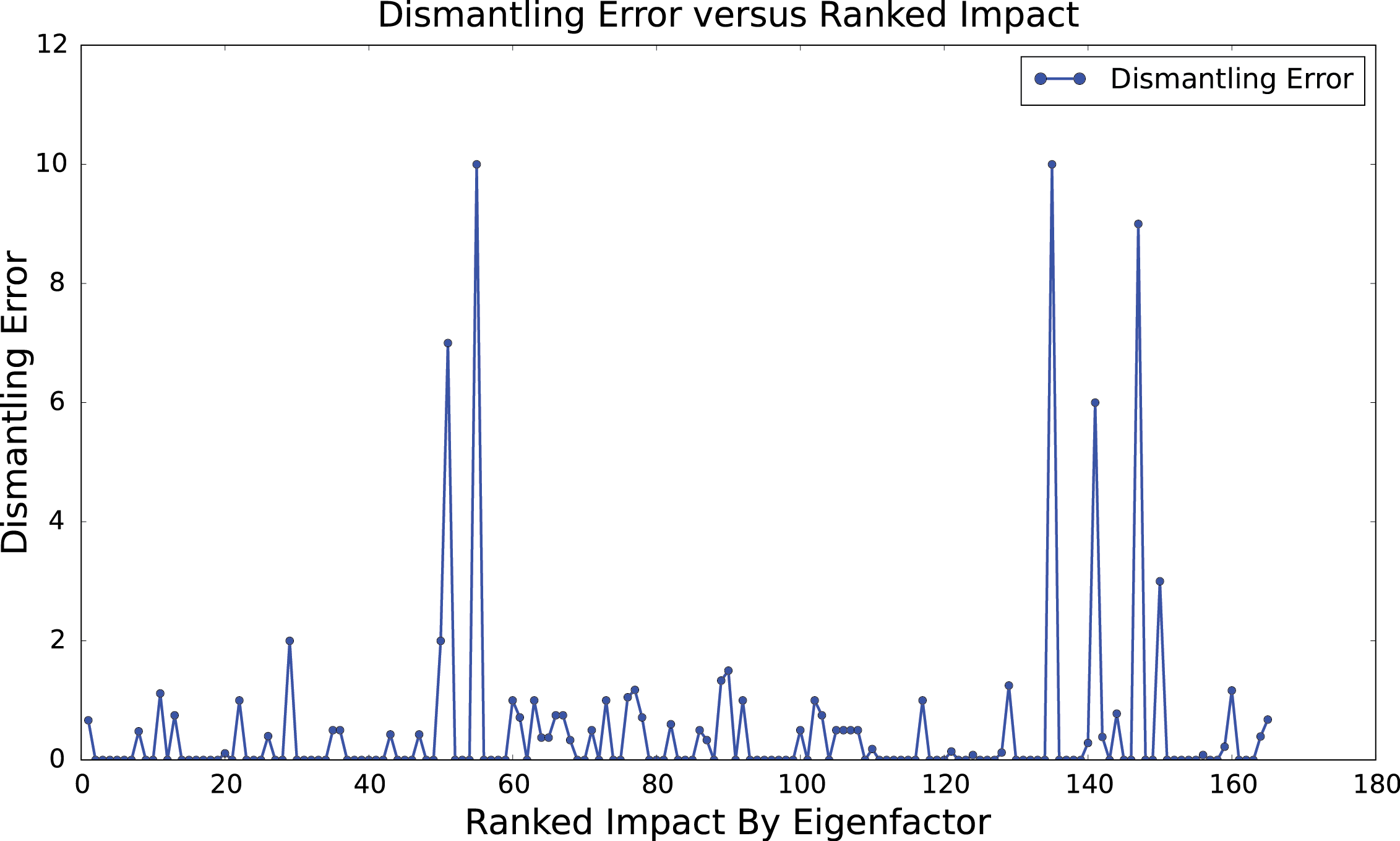}
 \caption{We randomly sampled 500 figures that are classified as multi-chart and compared their segmentation results to the ground-truth data (decomposed by human). We calculated the dismantling errors by calculating L1 norm of correct sub-figures and extracted sub-figures in each category. Then normalized the value to the number of correct sub-figures. It shows no correlation between the dismantling error and ALEF score of the source paper of the figure, eliminating one possible alternative explanation of our correlation result.}
 \label{fig:dismantler_bias}
\end{figure}

Table \ref{table:fig_stats} shows 67\% of figures are embedded in multi-chart figures. Our result could be explained by dismantling errors: if for high-impact papers the dismantler is more likely to generate broken fragments that are classified as diagrams, then that would explain our finding. In this section, we show that this explanation does not hold, as the dismantling errors are not biased with respect to the impact (Figure \ref{fig:dismantler_bias}).

We randomly sampled 500 figures that are classified as multi-chart and review their dismantling results. We manually labelled the sub-images into 8 categories: equation, diagram, photo, table, plot, fragment, multi-chart, and composite, where a fragment is a sub-image containing only missing text. We also manually generated the ground-truth data of ideal dismantling and counted the number of figures in each categories (no fragment, multi-chart, and composite in this case). We use the criterion proposed in our previous work \cite{Lee2015} that considers an array of photographic images to be one unit if the author assigns a part label for the array. For the case that the author assigns part labels for every photographic image, we consider them as independent photos to ensure that we do not artificially improve our results. The dismantler correctly extracted 82.9\% of the sub-figures from the 500 multi-chart figures and 84.3\% of the extracted sub-images are considered correct (not fragments, multi-charts and composites). We obtain a better result compared to our previous work because the testing images used in our previous work are all composite figures (comprising two or more types of figure). Correctly decomposing composite figures is usually more difficult then decomposing multi-chart figures with single figure type due to the higher possibility of unorganized layout found in composite figures. Figure \ref{fig:dismantler_bias} shows no correlation between dismantling error and article impact, where the dismantling error is mapped by 

\begin{equation*}
\frac{\sum_{i \in \textsf{categories}} \left | N^{\textsf{correct sub-figures}}_i - N^{\textsf{extracted sub-figures}}_i \right |}{\sum_{i \in \textsf{categories}} \left | N^{\textsf{correct sub-figures}}_i  \right |}
\end{equation*}

, where $N$ denotes the number of sub-figures. 

\begin{figure}[tb]
 \includegraphics[width=3.3in]{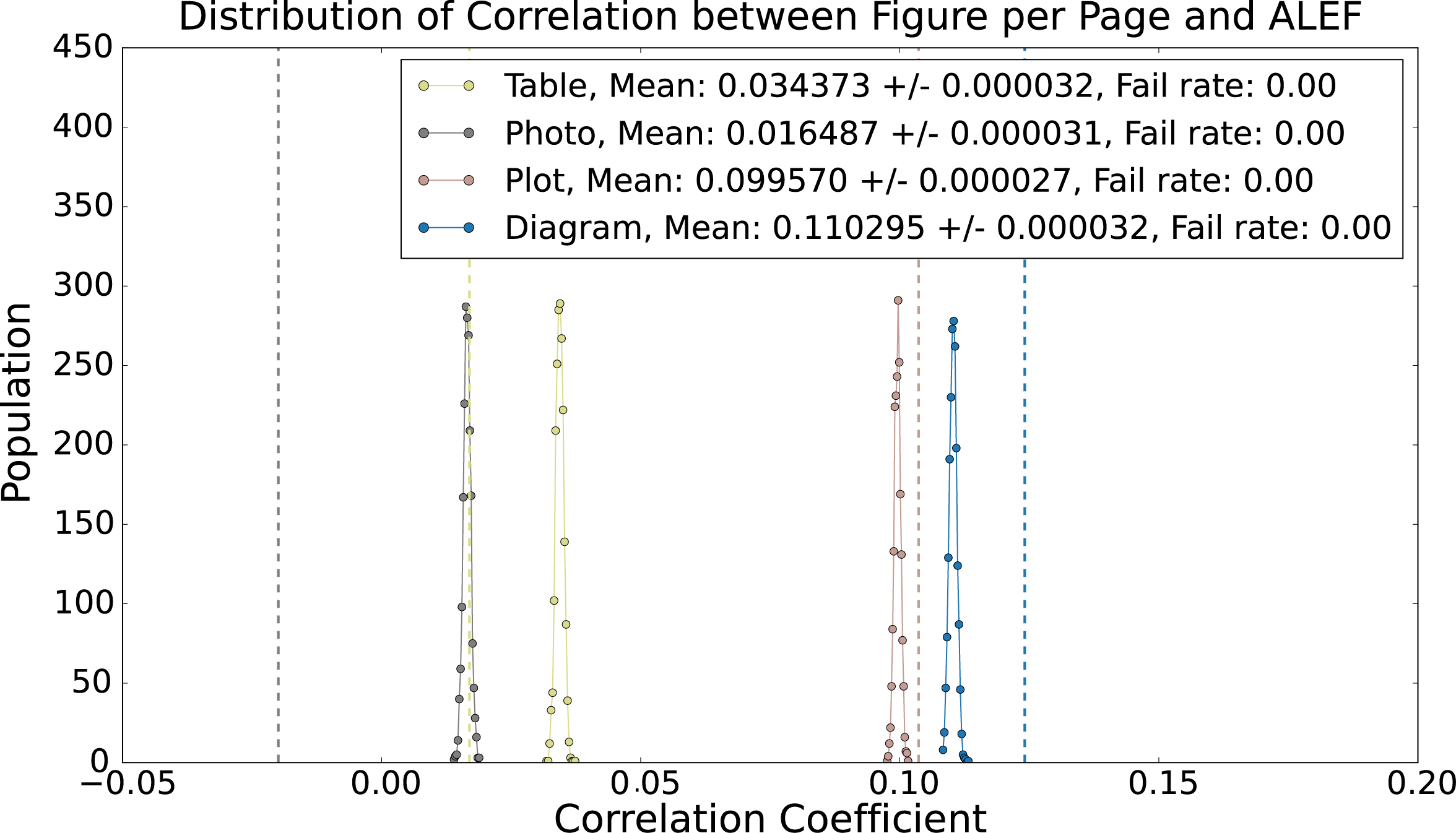}
 \caption{Calibrated correlation coefficients. Considering that the machine-labels can be mistaken, we calibrate the number of figures in each figure type by shuffling the machine labels according to the probabilities derived from Table \ref{table:confusion}. We performed 2000 trials to and plotted the distribution of the 2000 correlation coefficients. The dotted lines are the correlation coefficients obtained from the raw data without calibrating. The fail rate means the proportion of calibration experiments that produce non-significant correlations. The peak shift depicts the calibration effect and the peak width indicates the interval of possible correlation coefficients. From the calibration experiment, we obtained remarkable statistically significant correlations for plot (0.099570 +/- 0.000027) and diagram (0.110295 +/- 0.000032)
 }
 \label{fig:cor_distribution}
\end{figure}

\subsection{Classification Error Does not Explain Correlation Results}
\label{supl:error_impact}

In this section, we determine that the detected correlation between the article impact and figure use cannot be explained as a side effect of the errors of our classifier.

To adjust for this error, we correct the estimated counts of each figure type up or down by the known error rates for the classifier.  For example, if the classifier is known to misidentify a visualization as a diagram 10\% of the time, then we should adjust the estimated number of diagrams down by 10\% and the estimated number of visualizations up by 10\%.  We apply this correction for each cell in the confusion matrix produced from our evaluation.

After making this correction, the correlation coefficient is lower for diagrams and visualizations, but still well above zero (Figure \ref{fig:cor_distribution}).  However, this correction assumes that the errors are fixed; it is still possible that our classifier made an unusually bad guess and mislabeled many images as diagrams, generating a false signal.  To measure the likelihood of this case, we ran a series of 2000 experiments in which we randomly assigned a label based on the confusion matrix of the classifier.  For example, a figure that was originally labelled as a diagram may be relabeled as as an equation with a probability of 1.5\% (16/1057); a photo, 2.9\% (31/1057); a table, 5.5\% (58/1057); or a visualization, 12.2\% (129/1057). Otherwise, it remains a diagram.  By shuffling the labels this way, we can determine the sample distribution of our noisy classifier.  If the resulting distribution contains zero correlation within a 95\% confidence interval, then the signal we detected can be explained as a side effect of the errors of our classifier.

Figure \ref{fig:cor_distribution} shows the distribution of the unbinned correlation coefficients from the 2000 trials. The distributions for diagrams (blue) and visualizations (red) are still significantly above zero after the correction. The dotted lines are the correlation coefficients obtained from the raw data without corrections applied.  The peak shift depicts the correction effect and the peak width indicates the interval of possible correlation coefficients. From this experiment, we obtained remarkable statistically significant correlations for plot (0.099570 +/- 0.000027) and diagram (0.110295 +/- 0.000032).

\def\arraystretch{2.0}
\begin{table*}[tb]
\centering
\caption{Evaluation of figure-type classifier in consideration of false-positive singleton figures. This table shows the confusion matrix of the five categories. The numbers inside parentheses denote singleton figures, while those outside the parentheses denote the sum of singleton figures and multi-chart figures with single figure types. The multi-chart figures that comprise of two or more types of figure are defined as ``composite''. These figures are false-positive singleton figures. The ``Precision All'' considers only singleton figures as true positive and the denominators are 1400. Composite figures and multi-chart visualizations cause the low ``Precision All'' of diagram and failing of identifying photo arrays as multi-charts results the low precision of photo. About the ``Precision Singleton'', we eliminate all multi-charts and the values are more comparable with Table \ref{table:recall_precision} because singleton figures are the majority of our training set. We use this confusion matrix to derive the possibilities of inflation on the number of figures. These possibilities will be used to calibrate our raw data (see Supplementary Section: \nameref{supl:exp_ver}).}
\medskip
\label{table:confusion}
\begin{tabular}{l|c|c|c|c|c|c}
 	&Equation & Diagram & Photo & Table & Plot & Total\\
 \hline\hline
Equation & 1391(1391) & 16(16)  & 12(12) & 6(6) & 31(31) & 1456(1456) \\
Diagram & 4(4) & 850(823)  & 82(72) & 48(44) & 79(75) & 1063(1018)\\ 
Photo & 2(2) & 62(31)  & 1205(644) & 5(2) & 37(28) & 1311(707)\\
Table & 0(0) & 58(58)  & 0(0) & 1265(1263) & 9(9) & 1332(1330)\\
Plot & 3(2)  & 331(129) & 32(24)& 47(36) & 1195(1088) & 1608(1279)\\
Composite & 0 & 83 & 69 & 29 & 49 & 230 \\
\hline
Total & 1400(1399) & 1400(1057) & 1400(752) & 1400(1351) & 1400(1231) & 7000(5209) \\
\hline
Precision All & 99.4\% & 58.8\% & 46.0\% & 90.2\% & 77.7\%  & 74.4\%\\ 
Precision Singleton& 99.4\% & 77.9\% & 85.6\% & 93.5\% & 88.4\% & 90.0\%\\ 

\end{tabular}
\end{table*}

\end{document}